\def\year{2021}\relax
\documentclass[letterpaper]{article} %
\usepackage{aaai21}  %
\usepackage{times}  %
\usepackage{helvet} %
\usepackage{courier}  %
\usepackage[hyphens]{url}  %
\usepackage{graphicx} %
\usepackage{natbib}  %
\usepackage{caption} %
\nocopyright
\newcounter{rq}[section]
\newenvironment{rqenv}[1][]{\refstepcounter{rq}
   \textbf{RQ\therq: #1}}{}

\usepackage{array}
\newcolumntype{x}[1]{>{\centering\arraybackslash\hspace{0pt}}p{#1}}

\title{Teaching Tech to Talk: K-12 Conversational Artificial Intelligence\\ Literacy Curriculum and Development Tools\footnote{This work was funded by the Amazon Future Engineers program and Hong Kong Jockey Club Charities Trust.}}

\author{
    Jessica Van Brummelen,
    Tommy Heng,
    Viktoriya Tabunshchyk
}
\affiliations{
    \\Massachusetts Institute of Technology\\
    Cambridge, MA\\
    jess@csail.mit.edu, theng@mit.edu, vikt@mit.edu
}

\begin{document}

\maketitle

\begin{abstract}
With children talking to smart-speakers, smart-phones and even smart-microwaves daily, it is increasingly important to educate students on how these agents work---from underlying mechanisms to societal implications. Researchers are developing tools and curriculum to teach K-12 students broadly about artificial intelligence (AI); however, few studies have evaluated these tools with respect to AI-specific learning outcomes, and even fewer have addressed student learning about AI-based conversational agents. We evaluate our \emph{Conversational Agent Interface for MIT App Inventor} and workshop curriculum with respect to 8 AI competencies from the literature. Furthermore, we analyze teacher ($n$=9) and student ($n$=47) feedback from %
workshops with the interface and recommend that future work leverages design considerations from the literature to optimize engagement, collaborates with teachers, %
and addresses a range of student abilities through pacing and opportunities for extension. We found students struggled most with the concepts of AI ethics and learning, and recommend emphasizing these topics when teaching.
\end{abstract}

\section{Introduction}

Artificial intelligence (AI) literacy is becoming increasingly important in a world where what we see, hear and learn is often dictated by algorithms. Children decide what to watch based on AI recommendations; they talk to Siri or Alexa for help with math homework; they use online path planning algorithms to navigate to friends' houses. Technology's inner workings are often masked by %
simple user interfaces, and few users truly know how computers provide them with the information they receive \citep{folk-theories}.

Recent calls to action to create tools and curriculum to teach K-12 students about AI have produced easy-to-use classification model development tools, like Teachable Machine \citep{teachable-machine} and Machine Learning for Kids \citep{ml4kids}; day- to year-long AI curricula \citep{one-year-curriculum,smileycluster}; and interactive activities \citep{gesture-ml-abbie,popbots} to teach students about AI. However, very few of these have been analyzed with respect to AI literacy frameworks to determine how well they teach students particular competencies. This is largely due to the nascency of the K-12 AI education field: Researchers have only recently developed relevant AI literacy frameworks, like the Big AI Ideas \citep{big5}, AI Competencies and Design Considerations \citep{AILiteracy}, the Machine Learning Education Framework \citep{ml-framework-natalie} and AI extensions to computational thinking (CT) frameworks \citep{vanbrummelen-ct-ai}.

One recent work which used the AI literacy framework \citep{AILiteracy} to develop its curriculum, teaches AI competencies with Teachable Machine \citep{k12ml}. Nevertheless, it has not been %
analyzed for student learning outcomes. %
One work that does assess student learning of AI competencies involves teaching linear regression and gradient descent under three different conditions \citep{personal-data-ml}. %
However, this activity is for undergraduate students rather than K-12. Other works using the Big AI Ideas as frameworks to structure K-12 curricula include reinforcement learning activities in \emph{Snap!} \citep{rl-snap}, AI ethics curriculum \citep{ai-ethics-blakeley} and Zhorai \citep{zhorai}, but very few seem to directly assess student understanding of particular ideas.

In this work, we build on K-12 AI curriculum from \citep{vanbrummelen-sm}, in which students develop conversational agents using an interface in MIT App Inventor \citep{appinv}. We add presentations, interactive activities and student assessments developed according to \citeauthor{AILiteracy}'s AI literacy design recommendations and competencies. %
Students are assessed on eight competencies from the AI literacy framework.

Through feedback from two week-long workshops, we investigate two main research questions:

\begin{rqenv}\label{RQ:learning-ai}
    How does building and learning about conversational agents affect students' understanding of AI and conversational AI competencies?
\end{rqenv}

\begin{rqenv}\label{RQ:teaching-ai}
    What are effective teaching methods and curriculum content for AI literacy workshops?
\end{rqenv}

To address these questions, we present the conversational agent development interface, AI curriculum, results from assessments/feedback from 9 teachers and 47 students, and recommendations for future AI literacy tools and curricula.

\section{Motivational Scenario}
To provide a basis for understanding the conversational agent development interface, %
we present a scenario about how ``Sheila'' created a conversational AI cookbook app. Although the scenario is fictional, the app was based on a student's final project in the pilot workshop.%

\subsection{Sheila's Cookbook App}
\begin{figure}[htb!]
    \centering
    \includegraphics[width=\linewidth]{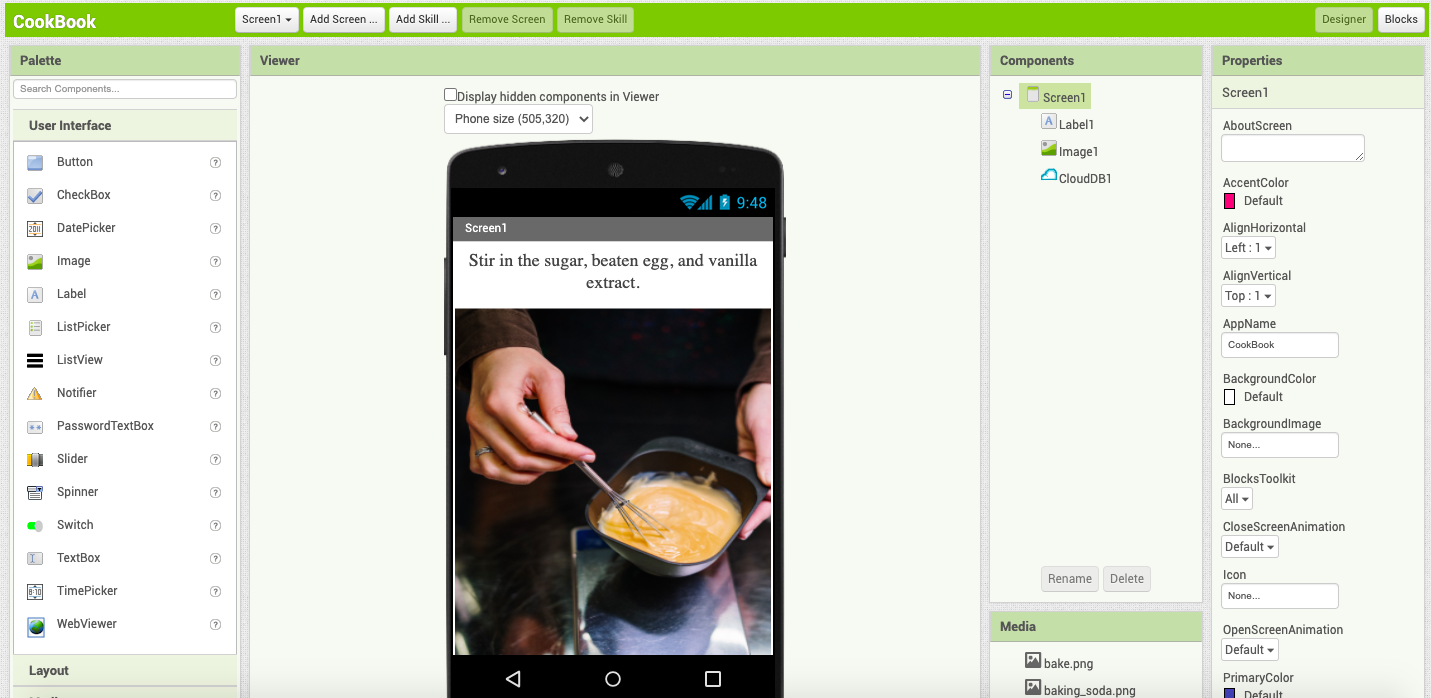}
    \caption{The cookbook mobile app being developed on the MIT App Inventor website.}
    \label{fig:app}
\end{figure}

Sheila, a ninth grade student, has recently found a passion for cooking. She has enjoyed trying different recipes and sharing with her family, but finds it hard to follow instructions on her tablet when her hands are messy. %
During a computer lesson, she heard about an experimental interface for developing conversational agents with MIT App Inventor and had a brilliant idea: to create a talking cookbook. Sheila would create a recipe app for her tablet, and enable conversation with Amazon Alexa. %
She would be able to ask Alexa for the ingredients or the next step in the recipe, and the app would display a relevant image %
on-screen (see Fig. \ref{fig:app}).

To implement the cookbook app, Sheila first found relevant pictures, %
uploaded them to MIT App Inventor, and stored recipe information in \emph{dictionaries} as \emph{global variables}. %
Sheila then added a \emph{CloudDB} component to her app, %
facilitating communication between the app and the Alexa skill (or voice-based app for Alexa) she would create. Finally, she added an \emph{event} block, which would update the app's screen when Alexa spoke about the next step in the recipe. %

After developing her mobile app, %
Sheila moved to the conversational AI portion of the project. When she added an Alexa skill to her project, special conversational AI code-blocks became available. 
First, she dragged out a \emph{define slot} block (see Fig. \ref{fig:slot}) and set the slot type to \emph{Food} so that the Voice User Interface (VUI) would recognize when someone mentioned food. %
She then dragged in three \emph{define intent} blocks %
to specify various ways users may interact with her skill: (1) ask for ingredients, (2) ask for the first step and (3) ask for the next step. She had heard that by giving Alexa \emph{training example} sentences for each intent, Alexa would \emph{learn} to categorize new sentences too. Finally, Sheila defined the skill's \emph{endpoint} function, %
being sure to include \emph{CloudDB} blocks to enable communication with her app. %

When Sheila was satisfied with her skill's blocks, she sent the code to Amazon by logging into her Amazon Developer account in App Inventor, %
and tested it while baking lemon scones. She was thrilled with how much easier it was not worrying about sticky hands on her tablet! She was also excited to see how Alexa understood her intent, even when not using the exact phrases she coded into her skill earlier. As she had heard in class, that was \emph{transfer learning} at work!

\section{AI Literacy Workshop Curriculum Design}\label{sec:curriculum}

We designed the curriculum for a remote workshop series running five days, 2.5 hours per day over Zoom for students in 8-12th grade with little-to-no experience programming. The only requirement was an internet-connected computer and the ability to test Android apps and Alexa skills, which could be simulated on the computer. %
Each day except the first concluded with a questionnaire. %
All tutorials were taught step-by-step. The curriculum addressed specific AI competencies from %
\citealt{AILiteracy} (see Tab. \ref{tab:competencies}).%

\begin{table}[htb]
\fontsize{9.0pt}{10.0pt}\selectfont %
\begin{tabular}{ll|ll}
\# & Competency                   & \# & Competency                   \\ \hline
1  & Recognizing AI               & 10 & Human Role in AI             \\
2  & Understanding Intelligence   & 11 & Data Literacy                \\
3  & Interdisciplinarity          & 12 & Learning from Data           \\
4  & General vs. Narrow           & 13 & Critically Interpreting      \\
5  & AI Strengths \& Weaknesses &    & Data                         \\
6  & Imagine Future AI            & 14 & Action \& Reaction           \\
7  & Representations              & 15 & Sensors                      \\
8  & Decision-Making              & 16 & Ethics                       \\
9  & ML Steps                     &  17 & Programmability              
\end{tabular}
\caption{\small List of AI Competencies from \protect{\citealt{AILiteracy}}.}\label{tab:competencies}
\end{table}

\textbf{Day 1.} The first day built students' familiarity with App Inventor and programming concepts relevant to conversational AI, such as variables, control statements, and events, through app-building tutorials. %
The first app counted and displayed the number of times the end-user pressed a button. The second app was a rule-based conversational agent, which would check if the text typed in a text-box was equal to ``Hello'' and respond with either ``Howdy'' or ``I don't understand''. %
Afterwards, students were encouraged to expand the phrases which the agent understood, introducing them to the difficulties of developing rule-based agents (providing a segue into developing machine learning-based agents).

\textbf{Day 2.} Next, we taught students about AI and conversational agents through presentations on the Big 5 AI Ideas \citep{big5}, which focused on Competency 1, 2, 3, 7, 8, 9, 11, 12, 14, 15 and 16 \citep{AILiteracy}, and AI ethics. The Big 5 AI Ideas presentation concluded with an interactive activity where students discussed whether different items (e.g., automatic door) might employ AI or not \citep{ai-or-not}, addressing Competency 1.  %
During the ethics presentation, students discussed bias with respect to conversational AI (e.g., sexist speech recognition due to homogeneous datasets), ethics of hyper-realistic text/speech generation, and the strengths and weaknesses of AI with respect to %
jobs, addressing Competency 5 and 16.

Finally, we introduced students to the experimental \emph{Conversational Agent Interface for MIT App Inventor} and key conversational AI vocabulary (e.g., invocation name, intent, utterance) through a tutorial. %
This and following tutorials used the interface to address Competency 10 %
and Competency 17 %
through experience programming conversational agents. Afterwards, students completed a questionnaire assessing their understanding of the Big 5 AI Ideas \citep{big5}, specifically focusing on Competency 1. %

\textbf{Day 3.} The next day, students learned how machine learning (ML) plays a role in Alexa skills through discussing the difference between the rule-based AI (Day 1) and Alexa skill (Day 2) tutorials. We also presented on transfer learning, and the feed-forward/back-propagation ML steps.  %
The presentations emphasized Competency 7, 8, 9, 11 and 12. Additionally, students completed the \emph{MyCalculator} tutorial, which demonstrated the fundamental conversational AI concept of extracting information from utterances using \emph{slots}. %

Finally, we taught students about different ML architectures (e.g., fully-connected networks, LSTMs) %
and engaged students in interactive demonstrations with particular models (e.g., GPT-2 text generation with \citealt{aidungeon-demo}). %
This demonstration and discussion focused on Competency 5 %
and 6. %
Afterwards, we asked students to contrast rule-based AI from ML-based AI %
to assess Competency 8. %

\textbf{Day 4.} Next, we taught students how to program communication links between mobile apps and Alexa skills with the \emph{ReadTheText} tutorial. This tutorial uses the \emph{CloudDB} component %
to enable communication between apps and skills \citep{clouddb-natalie}. We taught data literacy and representation (Competency 11 and 7) through the concept of the cloud.

We concluded Day 4 with a brainstorming activity on \emph{Padlet} \citep{padlet}, in which students %
contributed ideas for conversational agent final projects and thought about Future AI (Competency 6). %
Finally, they completed a questionnaire about transfer learning and generalization of intent phrases to assess understanding of Competency 12. %

\textbf{Day 5.} On the last day we connected with and supported students as they developed their final project. Students entered virtual breakout rooms with a ratio of instructors to students of approximately 2:8 (there were at least two adults in each room). Students created slides outlining their project and could volunteer to present to the entire cohort. %
Afterwards, students completed a final %
questionnaire, which asked them about their perception and understanding of AI.

\section{Design and Technical Implementation}

\begin{figure}[htb!]
    \centering
    \includegraphics[width=\linewidth]{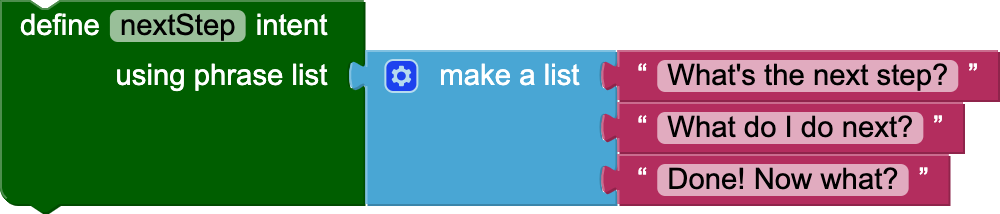}
    \caption{The \emph{define intent using phrase list} block. Each pink block represents a possible utterance for the intent.}
    \label{fig:utterances}
\end{figure}

Information about the design and implementation of the conversational agent interface can be found in \citealt{vanbrummelen-sm}. %
To summarize, the interface empowers students to learn CT/AI skills while developing agents that can converse, %
share data with mobile phone apps, %
and generate responses using ML. %
We draw extra attention to two blocks especially relevant to conversational AI competencies:

\textbf{Define intent using phrase list (Fig. \ref{fig:utterances}):}
In the \emph{define intent} block, students can enumerate utterances users might say to trigger each intent. When testing their skills, they were encouraged to try saying slightly different utterances from those they had enumerated (e.g., if they enumerated ``hello'' and ``hi'', they might say ``howdy'' instead), to see how ML-based systems can generalize over intent meaning, 
as opposed to a rule-based approach where only pre-programmed, exact utterances would be matched to an intent. This block plays a key role in teaching Competency 12. %

\begin{figure}[htb!]
    \centering
    \includegraphics[width=\linewidth]{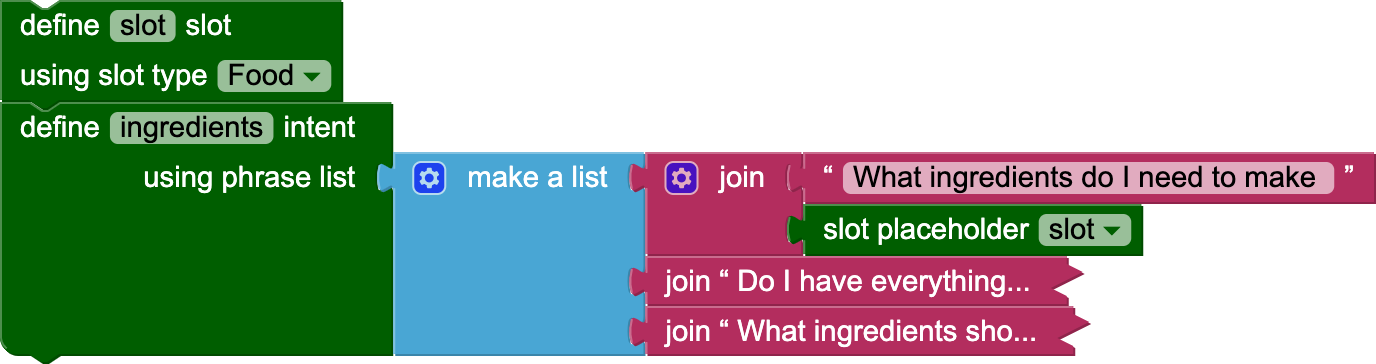}
    \caption{Slot blocks. Users can define slots of specific types, and use them when defining intents.}
    \label{fig:slot}
\end{figure}

\textbf{Define slot using slot type (Fig. \ref{fig:slot})}: 
Students can use \emph{slot} blocks to increase intent utterance flexibility. Slots act as placeholders for words end-users say to Alexa. For example, a ``food'' slot may be filled by saying the word, ``pizza''. This block teaches Competency 7 %
by encouraging students to think about how %
data is represented through slots.%

\textbf{Testing Alexa skills:} After programming skills, students could test them on any Alexa device, including the Android or iOS Alexa apps, or directly within the MIT App Inventor interface using a chatbot-like user interface (UI).

\section{Methods}
\subsection{Pilot Study}
We conducted a small pilot study with 12 students (7 girls, 3 boys, 2 did not answer; %
grade range 6-11, M=8.42, SD=1.56) voluntarily recruited by 2 teachers (with 12 and 16 years experience teaching) to test the interface and %
workshop content. During the pilot study, %
the text generation block \citep{vanbrummelen-sm} %
could not handle the high load of simultaneous users, so we replaced the activity that used the block %
with a text generation role-playing game activity \citep{aidungeon-demo}. We also streamlined an uploading process in the first tutorial, to reduce student confusion in the full study. 
Finally, students in the pilot were given Echo Dots, whereas in the full study, they were not.  %
Students in both studies could still speak to Alexa using the Alexa app or online simulator.

\subsection{Full Study}
\subsubsection{Participants}
Thirty-five students participated in the study (18 girls, 13 boys, 4 did not answer; %
grade range 6-12, M=9.97, SD=1.83), voluntarily recruited by 7 teachers (with 5-32 years experience teaching, M=14, SD=9.27) %
that signed up through an Amazon Future Engineers call for teachers at Title I schools. %

\subsubsection{Procedure} 
The experiment was conducted over 2.5-hour long sessions with students and teachers, followed by a half hour debrief with only teachers for 5 consecutive days. Teachers, parents and students completed IRB approved consent/assent forms prior to the workshops. %
Throughout the week, students connected to the workshops via Zoom. %
Students were given three daily questionnaires to fill out after class and three questionnaires to fill out during the workshops. Daily questionnaires averaged at 18 responses, and the in-class Day 1, Day 2, and Post-Workshop questionnaires had 33, 31, and 27 responses respectively.

\subsubsection{Data Analysis}
The dataset from the pilot and full-study workshops included student-made slidedecks describing their projects, screenshots of Padlet \citep{padlet} brainstorming sessions %
(as shown in \citealt{gist-appendix}), and quantitative and free-form answers from teacher and student questionnaires. (Note that not all of the results from the questionnaires are reported in this paper, due to space constraints and irrelevancy.)%

Three researchers performed a reflexive, open-coding approach to thematic analysis \citep{thematic-analysis-open-coding} for the free-form questions, typically inductively developing codes, and also drawing on literature, like the Big AI Ideas \citep{big5}, where appropriate. After each researcher separately completed the familiarization and generating-code stages, and several discussions, we came to consensus on codes for each of the questions. %
Codes and representative quotations can be found in \citealt{gist-appendix}. (Note that some answers were tagged with multiple codes as they encompassed multiple ideas.) For questions asked on both pre- and post-questionnaires, we used the Wilcoxon Signed-Rank Test to assess learning gains. Additionally, we report results from demographics and AI literacy competency multiple choice questions.

\section{Results}
We divided the results into two sections based on our research questions, which address competencies students learned about AI and conversational AI (RQ\ref{RQ:learning-ai}) and the effectiveness of how our workshops were taught (RQ\ref{RQ:teaching-ai}).

\subsection{RQ\ref{RQ:learning-ai}}
To address whether students learned specific AI competencies from \citealt{AILiteracy} (see Tab. \ref{tab:competencies}) and conversational AI competencies, we analyzed answers to assessment questions. %
The majority of the questions were free-form and were assigned between 2 and 13 different codes.%

\subsubsection{Knowledge Gains}
In both the pilot and full study, most students had not developed a conversational agent before (90\% in each case). To assess knowledge gains in terms of understanding AI and decision-making, ({Competency 2 and 8) as well as conversational AI, we asked students questions prior to and after the workshops. For understanding AI, we looked for answers recognizing one or more of the Big AI Ideas \citep{big5}; for conversational AI, we looked for answers involving systems understanding natural language (NL) and responding in NL, which were the two key components of conversational agents we described in the Day 2 presentation. %
To assess decision-making, we asked students how conversational agents decide what to say. 

For both competencies, the data indicated a better understanding after the workshops. (This is despite accidental priming of some students by answering questions about AI prior to the first questionnaire.) Students cited the Big AI Ideas more frequently in the post- vs. pre-questionnaire for both Competency 2 (40\% vs. 21\%, $Z=2.27$, $p=0.02$) and Competency 8 (41\% vs. 21\% for Competency 8, $Z=2.22$, $p=0.03$). One student showed particularly good understanding of the ML steps ({Competency 9) after the workshops, describing how conversational agents are ``trained by being shown lots of samples'', engage in ``testing of these skills'', and use ``complex and fast math to derive an output from an input''. %
In terms of understanding conversational AI, we found no evidence for significant difference between the number of tagged answers alluding to both a NL understanding and response in the post- (43\%) vs. pre-questionnaire (29\%), as well as answers alluding to at least one of understanding or response (53\% vs. 59\%).

\subsubsection{Other Competency Assessments}
For the second day's homework question, we asked students about how they might investigate whether a ``smart-fridge'' truly utilized AI and conversational AI to assess Competency 1. %
When investigating AI, the vast majority of tagged answers (86\%) corresponded to Big AI Ideas \citep{big5}. Only 14\% were shallow or unrelated to the Big Ideas. When investigating whether the fridge was conversational AI, a majority of answers (67\%) alluded to at least one of NL understanding or response, and 65\% of those alluded to both understanding and response. Twenty percent showed understanding of AI (e.g., Big AI Ideas), but not necessarily conversational AI, and only 13\% showed little understanding of AI at all.

Students were also asked to imagine future AI (Competency 6), by designing the next version of Alexa. The vast majority of students came up with %
ideas to improve Alexa devices, with only 4\% of tagged answers being vague or shallow. The most common %
ideas were related to the Big Idea of natural interaction (e.g., emotion recognition, improved speech recognition) (51\%) or adding a new/futuristic feature to the AI device (27\%). A significant number of them (18\%) had to do with societal implications (e.g., improving privacy/security, increasing the number of languages).

To assess Competency 5, %
we asked students what some of the positive and negative effects of creating a conversational agent to help their friend with math homework. Students cited a variety of reasons for the strengths of AI (71\% of tags), including constant availability, personalized learning, and time efficiency. Nonetheless, 29\% of tagged answers seemed vague or shallow. For the weaknesses of AI, students seemed to more easily come up with answers, with only 6\% of tagged answers seeming vague or shallow. Students described how AI systems are more rigid, are less likely to understand speech, and can create ethical problems. %

In addition to assessing %
Competency 8 through pre- and post-questions (outlined above), %
we asked students to describe differences between rule- and ML-based AI. Students seemed to understand both ML-based AI (with 50\% of tags alluding to the Big Ideas of learning and representation \& reasoning), and rule-based AI (with 48\% of tags alluding to the limitations and programmatic nature of rule-based AI). Only one (2\%) tagged answer seemed shallow or unrelated.

In the workshops, students embodied the human role of programming AI agents (Competency 10); thus, we asked them for decisions they made (and developers might make) when creating agents. Not surprisingly, none of the answers seemed vague or shallow. Students described decisions related to natural interaction, societal impact, learning, and how to program specific functions. One example described how developers must ensure ``device[s are] able to recognize a variety of voive[\emph{sic}] types (for example, male and female) to minimize biases'', demonstrating how humans make ethical AI decisions.%

To assess Competency 12 %
we provided a scenario in which an Alexa skill was developed with particular training data (i.e., intent phrases) and asked whether the system would likely recognize (i.e., be able to learn) a phrase similar to the given data, but not exactly the same. With the given information, the system would have likely been able to learn the new phrase; however, responses were split evenly (50\%) between if it would be recognized or not. The most common reason for it not being recognized was because the phrase did not exactly match the training data (40\% of tagged answers), and the most common reason for it being recognized was due to its similarity to the training data (33\%). %

The final competency we directly assessed was students' understanding of AI ethics (Competency 16). When asked whether it would be okay to use AI to generate a bedtime story during the ethics presentation, 91\% of respondents said yes. When asked whether it would be okay to generate a news article---although we did not administer a poll for this question---responses from the discussion seemed more varied, with students indicating articles should ``always [have] a disclaimer'' stating ``that [an] AI system created it''. %

We also posed a question in the last questionnaire about the implications of millions of people using the student's final project. There was a wide range in answers, including positive implications for mental and physical health, education and the environment, as well as negative implications for privacy/security, about overreliance on AI, and how content may be offensive to some people. Of the tagged quotes, 57\% related to positive effects and 37\% related to negative. %

\subsection{Student Projects}
Students each developed a conversational agent project based on ideas generated in a brainstorming session, as shown in \citealt{gist-appendix}. %
Twenty-nine ideas were generated in the pilot, and 41 in the full study. Ideas ranged from tracking carbon emissions with voice to creating haptic feedback for users living with deafness. Ultimately, students entered information about their projects---including project names, target users and example conversations---in a slidedeck. Two exemplary projects are shown in Fig. \ref{fig:projects}. Of the projects in the slidedeck, 29\% were educational-, 26\% were mental health-, 21\% were productivity-, 8\% were accessibility-, 8\% were physical health-, 5\% were entertainment-, and 3\% were environmental-related skills.

\begin{figure}[htb!]
    \centering
    \fbox{\includegraphics[width=\linewidth]{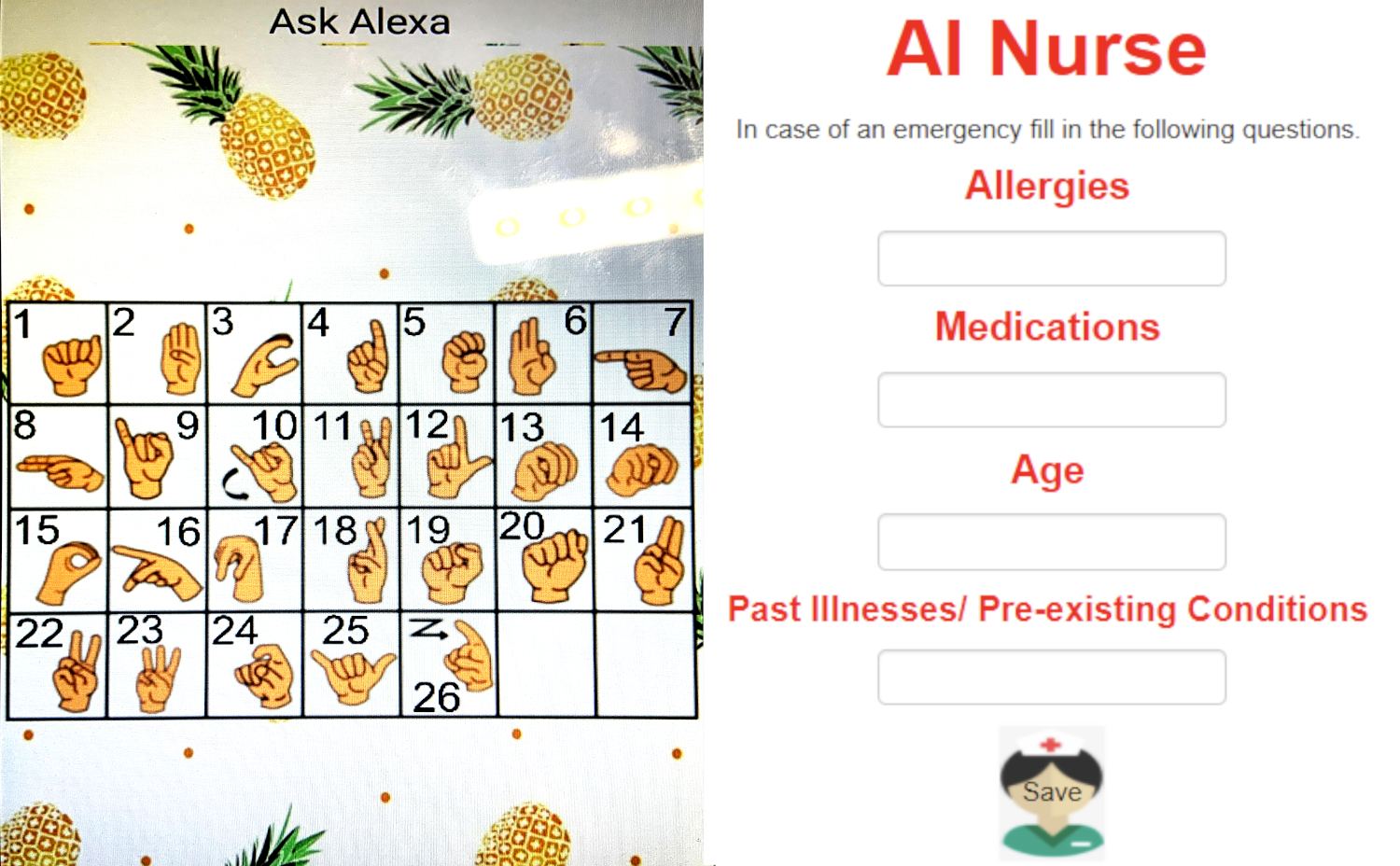}}
    \caption{Screens from apps for students' final projects. Each app communicated with Alexa skills: one of which helped users learn to sign, and the other, diagnose illnesses.}
    \label{fig:projects}
\end{figure}

\subsection{RQ\ref{RQ:teaching-ai}}
To address how we can teach conversational AI and AI competencies, we analyzed answers from teachers and students about engagement, interest, and content. %
Each free-form question was assigned 2-10 different codes and answers were tagged accordingly. %

\subsubsection{Teaching Method Effectiveness}
Overall, the results indicate the workshops were effective and useful for teachers and students. We asked teachers what they would leave behind or bring to their classrooms from the workshops. %
The only comment about not including something from the workshops 
had to do with giving each student an Alexa device (which we only did during the pilot), as this would not be feasible for an entire class. Seven of nine answers alluded to leaving the course ``as-is'' and/or to a desire to bring it into their classroom.%

We also asked teachers whether the material in the course changed their opinions about AI, conversational AI or teaching AI, and many teachers (5/9) reported that the course made understanding and teaching AI seem more feasible (e.g., ``teaching AI is daunting, however [the] materials presented and the step by step tutorials and explanations help[ed] me visualize how I can teach [...] AI in class''). %
Teachers also reported that their students were highly engaged (scale of 1-5, M=4.6, SD=0.497), making statements like, ``My students were 100\% engaged in the course. They loved coding and the fact that by the end of each day they had a working app made it especially interesting.'' The most frequently stated reason for engagement was the hands-on material---specifically the coding tutorials and group activities. One teacher mentioned that ``The only decline in engagement I noticed was due to pacing''. In the debrief sessions, one teacher mentioned how she thought some students might have been getting bored, since we typically waited until all students were caught up in the tutorial before continuing the lesson. In subsequent lessons, we sped up our pace and created breakout rooms in Zoom for students who were falling behind to catch up with the help of instructors.

We also asked students about the most interesting part of the class and their favorite overall activity. Answers related to making Alexa skills were most common (35\%), then AI-related answers (31\%), and group activity-related answers (13\%). %
Students' favorite part of the workshops was programming (38\%), including tutorials, final projects and seeing code working on devices, and then learning about AI (19\%). Students also mentioned their appreciation for the teachers' helpfulness (``I really enjoyed how kind the speakers were, and how they would always help if you needed it'') and pace (``Going slowly was helpful for learning'').

Additionally, we asked whether students enjoyed using the physical Alexa device (if they had one or were given one in the pilot; 41\% of respondents) or if they would have liked to try the skill on a device (if they were using the Alexa app or simulator; 59\%). Most responses noted enjoying seeing programs on-device (40\%), enjoying exploring device features (17\%) or wanting to see their programs on-device (33\%), with only 10\% reporting the device did not seem necessary or desired.

\subsubsection{Improvements}
To determine how to improve the workshops, we asked teachers for daily feedback in the debrief sessions. Their feedback included practical ways to better pace the workshops, address engagement, and foster student interest. For instance, %
to increase interaction during the tutorials, teachers noted how we could ask students to unmute themselves on Zoom and demonstrate the conversational agents they developed. This not only allowed us to ensure students were following along, but also created space for instructors and students to connect. Other feedback included making Zoom polls to instantly assess students' progress and help us with pacing, and asking students about their projects to help them self-identify where they could improve, rather than commenting directly on how we think they could improve. This daily teacher feedback was integrated into the workshops as they progressed.

After the workshops, we asked teachers what they would change or add, and although many of the answers did not specify anything, some teachers noted they would extend the workshops with additional content, including ``practice activities that students can do outside of the workshop'' or collaborative activities. We also asked what students struggled with. The most common (54\%) tag had to do with technical issues, like the ``initial set up'' or ``accidentally delet[ing]'' blocks. The next-most common (31\%) tag was about complex terminology, %
and a few (15\%) mentioned the slow pace.

We also directly asked students what they struggled with most and how we could improve, and most often they noted struggling with programming (33\%), like ``understanding what some blocks are used for'', and technical difficulties (22\%), like ``getting the emulator to work''. Nineteen percent of tags indicated there was nothing to improve.

\subsubsection{Evidence for Learning}
To investigate if teachers and students felt the workshops were useful for learning, we asked them to self-report learned skills and understanding. When asked if they gained a better understanding of conversational AI, all teachers responded positively. One teacher noted they ``gained understanding on how to teach AI through modeling and guided practice. With the help of tutorial lesson and the explanation of the facilitators, [they now] understand how AI work[s].'' When asked what the most important skills and ideas their students learned, they cited programming (64\%), AI concepts (21\%), and societal impact of AI (14\%). %

Teachers were also asked to summarize key skills and ideas students learned. Most frequently, teachers mentioned conversational AI concepts (40\%), then blocks-based programming (28\%), then CT (20\%), and finally project ideation (12\%). To encourage recognition of skills developed through process, teachers were also asked if their students demonstrated ``maker skills'', which include risk-taking, collaborating, connecting knowledge, and persisting \citep{maker-skills}---or as one student put it, skills about ``trying, failing, and learning from prior experience''. %
Teachers provided examples of students participating in discussions (29\%), presenting projects (29\%), helping each other (18\%), persisting despite difficulties (18\%) and asking questions (6\%). Students also self-reported demonstrating maker skills, including examples of speaking up (38\%), persistently debugging (32\%), helping others (11\%), and creating extensions to the in-class tutorials (12\%). Only 8\% of responses did not indicate any of the maker skills.

\section{Discussion}
This section identifies AI competencies (see Tab. \ref{tab:competencies}) students learned well and which were more difficult to learn through the curriculum. It also examines the effectiveness of our remote-learning AI literacy curriculum, specifically noting AI literacy design considerations \citep{AILiteracy} and identifying areas for improvement.

\textbf{Most AI competencies were learned well.}
As evidenced in the results section, most answers showed mastery of relevant AI literacy competencies from \citealt{AILiteracy}. Student feedback also showed excitement and interest in learning AI. For example, one student said they were ``impressed with [themself] and how much information [they were] able to retain from this workshop''; %
another said the most interesting thing they learned was ``what makes an AI an AI''. Our workshops directly assessed Competency 1, 2, 5, 6, 8, 10, 12 and 16. Future research should consider how to address all competencies. %
Furthermore, although the number of answers tagged with understanding the conversational AI competency increased from pre- to post-workshop, the difference was not significant. This may have been due to a larger focus on understanding AI generally rather than conversational AI in our presentations. Future workshops should consider increased focus on specific types of AI.

\textbf{Machine learning and ethics are difficult concepts.}
Students did not learn certain competencies as well as others. For example, 50\% of answers to the question addressing learning (Competency 12) did not indicate understanding of ML generalization. This may have been due to students being more familiar with rule-based systems (as 76\% of students had previous programming experience) that do not exhibit learning. Particular attention to this concept while teaching could enable students to better understand it in future workshops.

Another interesting result had to do with AI ethics (Competency 16) and AI strengths \& weaknesses (Competency 5). When addressing how people would be affected by their skill's audience growing to millions of users, students cited many more positive effects (57\%) than negative (37\%). While it may be true that there would be more positive effects, negative effects are critical to understanding the ethics of AI technology. Nonetheless, when answering the question about the positive and negative effects of developing a conversational agent to help a friend with math homework, students presented many more vague or shallow answers (29\%) for positive effects than negative (6\%). One reason for this discrepancy may be a bias towards seeing the positive in ones' own technology. Future workshops should focus on inadvertent implications of AI, perhaps by using an ethical technology framework like the one in \citealt{leung-sm}.

\textbf{Engaging teachers is vital to educational research success.}
Teachers' invaluable knowledge and experience in the classroom  %
helped us implement AI literacy design considerations \citep{AILiteracy}. For example, teachers noticed how
particular aspects of our UI could be improved to foster a low barrier to entry (Design Consideration 15), like %
streamlining the uploading process in the first tutorial. They also contributed practical in-class methods to increase engagement, like promoting social interaction during tutorials (Design Consideration 11) and leveraging learners' interests (Design Consideration 12) by asking students' opinions on projects, rather than directly commenting on them. Teachers' feedback was vital to our workshop's success.

Additionally, to democratize AI education, K-12 teachers need to be empowered to understand and teach AI. Through engaging in our study, teachers were able to better grasp AI concepts and how to teach AI. In the post-questionnaire, teachers mentioned how the workshops made AI more accessible and feasible to teach, and many of them emailed us afterwards asking if they could use the materials in their classrooms. %
Further teacher-researcher collaboration is encouraged to better develop AI education resources and bring these resources directly to classrooms. This idea is supported by other education research \cite{ed-innovation-codesign}; however, to the authors' knowledge, no studies have yet co-designed AI curriculum with teachers.

\textbf{Workshop pacing should meet all students' needs.} %
To encourage a low barrier to entry (Design Consideration 15), we paced the workshop such that no student fell behind. This was to avoid the pitfall observed in other workshops in which students felt rushed \cite{masters-natalie, vanbrummelen-sm}. Feedback on the pacing was mixed. Some teachers and students appreciated our ``patience'' and slow pace, as it was ``helpful for learning'', especially for ``the students who [typically] fall behind'', who ``are exactly the ones we need to bring along''.  %
Others, however, felt the pace reduced engagement. One comment from teachers encouraged us to provide written tutorials for advanced students such that they could go ahead in the tutorials on their own. We did so, and some students began to complete the tutorials early, adding new intents and other extensions to the conversational agents with the extra time. One student mentioned how they ``liked to go ahead [in the written tutorial] for a bit and re-listen to why specific coding was used''. Future AI workshops should pay specific attention to students' pacing needs, providing extension opportunities for advanced students, while still providing a low barrier to entry.%

\textbf{Hands-on, interactive activities, and leveraging learners' interests contributed to high engagement.}
According to their teachers, students were highly engaged throughout the workshops. Students found making Alexa skills, learning about AI, and group activities %
particularly interesting, which each embodied specific AI literacy considerations: making Alexa skills provided opportunities to program (Design Consideration 6); brainstorming personal projects leveraged learners’ interests (Design Consideration 12); and discussing AI ethics encouraged critical thinking (Design Consideration 8). %
These results are similar to other K-12 AI curricula \cite{pic-danny,vanbrummelen-sm}, which also implemented these design considerations.

\textbf{Physical devices were engaging, but not necessary.}
To contextualize conversational agent development, we gave students in the pilot Amazon Echo devices, which they used to test their Alexa skills. Students enjoyed using the physical device (e.g., because it felt ``more hands on'' and they got to ``see skills that [they] coded working in the same way that Amazons[\emph{sic}] skills work''); however, one teacher alluded to how providing each student with a physical device was not scalable---especially for Title I schools. Fortunately, the interface has a built-in chatbot UI for testing Alexa skills,%
so the workshops can be carried out without physical Alexa devices. Having alternatives to expensive hands-on technology is important for making AI accessible to everyone, and should be encouraged in future K-12 AI curricula.

\section{Limitations}
This study shows how a particular workshop curriculum addresses AI literacy. Further studies with larger sample sizes should be completed to confirm the effectiveness of the applied methods and generalize results to different contexts. Additionally, the results about knowledge gained may have been skewed due to researchers answering questions about AI prior to the pre-questionnaire (giving some students additional initial knowledge about AI). Results may have also been affected due to varied remote learning environments. %

\section{Conclusions and Future Work}
This paper presents AI literacy-focused curriculum to teach conversational AI concepts. Through interactive AI workshops, students learned AI competencies and developed conversational agents. We found evidence for the effectiveness of AI design consideration-based curriculum to engage students and teach AI competencies. We also identified competencies students had difficulty with (ML and ethics), which should be focused on in future work. The materials from this workshop and a demo video can be found in the appendix \citep{gist-appendix}. %

\section{Acknowledgments}
We thank the teachers and students who participated in our study, volunteer facilitators, MIT App Inventor team, Personal Robots Group, and Amazon Future Engineer members who made the workshops possible. Special thanks to Hal Abelson, Daniella DiPaola, Jennifer Leung, Cheri Ackerman and Evan Patton for comments on the manuscript; Karen Lang, Selim Tezel and Randi Williams for curriculum guidance; and Hilah Barbot for the AFE connection.

\fontsize{9.0pt}{10.0pt}\selectfont
\bibliography{paper}
\bibstyle{aaai}

\end{document}